# DIELECTRIC AND REFRACTIVE INDEX MEASUREMENTS FOR THE SYSTEMS 1-PENTANOL + 2,5,8,11,14-PENTAOXAPENTADECANE, OR FOR 2,5,8,11,14-PENTAOXAPENTADECANE + OCTANE AT (293.15-303.15) K.

VÍCTOR ALONSO, JUAN ANTONIO GONZÁLEZ*, ISAÍAS GARCÍA DE LA FUENTE, JOSÉ CARLOS COBOS

G.E.T.E.F., Departamento de Física Aplicada, Facultad de Ciencias, Universidad de Valladolid, 47071 Valladolid, Spain,

*e-mail: jagl@termo.uva.es; Fax: +34-983-423136; Tel: +34-983-423757

**ABSTRACT**

Relative permittivities, $\varepsilon_r$, and refractive indices, $n_D$, have been measured at (293.15-303.15) K, for the mixtures 1-pentanol + 2,5,8,11,14-pentaoxapentadecane (TEGDME) or TEGDME + octane. These data have been used, together with density measurements available in the literature, to calculate the correlation factors, $g_K$, according to the Kirkwood-Fröhlich equations. The curves of the deviations of $\varepsilon_r$ from the ideal behaviour, $\Delta\varepsilon_r$, of the 1-pentanol + TEGDME system are shifted to high mole fractions of the alcohol, and show a rather large minimum. The shape of the $\Delta\varepsilon_r$ curves of methanol + TEGDME or + polyethylene glycol dimethyl ether 250 (PEG-250) mixtures is similar. This reveals that polyethers can break the alcohol self-association even at high alkanol concentrations. The higher $\Delta\varepsilon_r$ values of the methanol solutions are ascribed to large self-association of this alcohol. The $\Delta\varepsilon_r$ curves of the 1-pentanol + dibutyl ether (DBE) system are nearly symmetrical, and the $\Delta\varepsilon_r$ values are higher than those of the corresponding TEGDME mixture. This indicates that effects related to the alcohol self-association are much more relevant in the DBE system. These findings are supported by $g_K$ and molar polarization data. Values of molar refraction of 1-pentanol systems reveal that dispersion forces become more important in the sequence: octane < DBE < TEGDME. Such forces are also more relevant in the 1-pentanol + TEGDME system than in the methanol + TEGDME solution. The comparison of $\varepsilon_r$ and $g_K$ data for TEGDME or DBE + octane mixtures shows than the polyether is a more structured liquid.

**KEYWORDS: permittivity; refractive index; 1-pentanol, TEGDME, correlation factor**

## 1. INTRODUCTION

1-Alkanol + ether mixtures are of high interest from both practical and theoretical points of view. For example, mixtures containing oxygenated compounds, such as ethers and/or alkanols are of great importance because they are increasingly used as additives to gasoline owing to their octane-enhancing and pollution-reducing properties [1,2]. Polyethers are important solvents in many chemical reactions such as Grignard reduction, or alkylation or organo-metallic reactions [3]. On the other hand, in a series of recent articles [4-7], we have shown some crucial features of 1-alkanol + ether systems. A short summary follows. (i) Effects related to self-association of 1-alkanols are determinant when considering the thermodynamic properties of mixtures with linear monoethers. (ii) Such effects are weakened if a linear monoether is replaced by a cyclic monoether, in such way that, from 1-hexanol, 1-alkanol + tetrahydrofuran, or + tetrahydropyran mixtures show a structure close to that of random mixing. (iii) The replacement of a linear monoether by a linear polyether also leads to a weakening of the mentioned effects relative to self-association of 1-alkanols, while dipolar interactions are increased. These results have been obtained from measurements on phase equilibria, molar excess enthalpies, molar excess heat capacities at constant pressure or from volumetric measurements and from the application of different theories such ERAS [8], DISQUAC [9], Flory [10], or the Kirkwood-Buff integrals formalism [11]. The investigation of the mixture structure can be also carried out on the basis of $\varepsilon_r$, and $n_D$, data. These magnitudes together with density data make possible the determination of the Kirkwood´s correlation factor, $g_K$, [12-15], which provides useful information on the mixture structure. As a part of a general systematic study in which liquid solutions are investigated using $g_K$, we report here $\varepsilon_r$ and $n_D$ measurements for the mixtures 1-pentanol + TEGDME, or TEGDME + octane at (293-15-303.15) K. Previously, we have investigated 1-pentanol + DBE, or + octane mixtures, or the DBE + octane system [16]. $\varepsilon_r$ and $n_D$ data at different temperatures for the methanol + TEGDME, or + PEG-250 systems [17-19], or for glyme + alkane mixtures [20-22] are available in the literature.

## 2. EXPERIMENTAL

*2.1 Materials*

1-Pentanol, octane and TEGDME were supplied by Fluka, and used without further purification. Their purity in mass fractions was ≥ 0.99; ≥ 0.99 and ≥ 0.98, respectively. Values of the physical properties of pure compounds, density, $\rho$, (measured with an Anton Paar DMA

602 vibrating-tube densimeter (uncertainty 5 g·cm$^{-3}$), thermostated within ± 0.01 K); $\varepsilon_r$, and $n_D$ are listed in Table 1. They are in good agreement with the literature values (Table 1).

*2.2 Apparatus and procedures*

Binary mixtures were prepared by mass in small flasks of about 10 cm$^3$. All weighings were corrected for buoyancy effects. The error on the final mole fraction is estimated to be lower than ± 0.0001. Conversion to molar quantities was done using the relative atomic mass Table of 1995 issued by I.U.P.A.C [23]. All the $\varepsilon_r$ and $n_D$ measurements were carried out under static conditions and atmospheric pressure

Permittivity measurements were taken using the Agilent 16452A cell, connected to a precision impedance analyzer model 4294A through a 16048G test lead, both also from Agilent. Temperature is controlled within ± 0.02 K by means of a thermostatic bath LAUDA RE304 and measurements were taken at 1 MHz. Details on the apparatus configuration and on the calibration procedure can be found elsewhere [16]. The $\varepsilon_r$ accuracy is of 3% or less. From the differences between our data and values available in the literature, the uncertainty of $\varepsilon_r$ is estimated to be l% or less.

Refractive indices were measured with a refractometer model RMF970 from Bellingham-Stanley. The temperature is controlled by Peltier modulus and the temperature stability is ± 0.05 K. Prior to the measurement at each temperature, the apparatus must be calibrated with a known reference, usually distilled and deionised water or toluene. The uncertainty of the $n_D$ measurements is 0.02% or better, as it is shown by the differences between our $n_D$ results and those reported in the literature for recommended liquids [24].

## 3. RESULTS

Table 2 lists, in the temperature range (293.15-303.15) K, values of $\varepsilon_r$ and of deviations of this magnitude from the ideal state, $\Delta \varepsilon_r$, vs. $x_1$, the mole fraction of the first component for 1-pentanol + TEGDME or for TEGDME + octane systems. For an ideal mixture at the same temperature and pressure than the system under study, the $\varepsilon_r^{id}$ values are calculated from the equation [22]:

$$\varepsilon_r^{id} = \phi_1 \varepsilon_{r1} + \phi_2 \varepsilon_{r2} \qquad (1)$$

where $\phi_i = x_i V_i / V^{id}$. Deviations from the ideal behaviour are calculated from the expression:

$$\Delta \varepsilon_r = \varepsilon_r - \phi_1 \varepsilon_{r1} - \phi_2 \varepsilon_{r2} \qquad (2)$$

Table 3 contains, for the same mixtures, $n_D$ values at (293.15-303.15) K, needed for $g_K$ calculations. Results on $\Delta \varepsilon_r$ are shown graphically in Figures 1-2. No data have been found in the literature for comparison. The $\Delta \varepsilon_r$ data have been fitted by unweighted least-squares polynomial regression to the equation:

$$\Delta \varepsilon_r = x_1(1-x_1)\sum_{i=0}^{k-1} A_i (2x_1 - 1)^i \qquad (3)$$

The number of coefficients used ($k$) in equation (3) for each mixture was determined by applying an F-test [25] at the 99.5 % confidence level. Table 4 lists the parameters $A_i$ obtained in the regression. The $n_D(x_1)$ measurements were fitted to the function:

$$n_D = \sum_{i=0}^{p} A_i x_1^i \qquad (4)$$

Values of the $A_i$ coefficients are given in Table 4, which also lists the corresponding standard deviations, $\sigma$, defined by:

$$\sigma(F) = \left[ \frac{1}{N-k} \sum \left( F_{cal} - F_{exp} \right)^2 \right]^{1/2} \qquad (5)$$

where $N$ is the number of direct experimental values and $F = \Delta \varepsilon_r$ or $n_D$. For the methanol + TEGDME, or + PEG-250 mixtures at 303.15 K, $\Delta \varepsilon_r$ values have been obtained in this work using $\varepsilon_r(x_1)$ measurements reported elsewhere [17,18]. The coefficients of the corresponding Redlich-Kister expansions were also determined. Results are shown graphically in Figure 1.

## 4.     DISCUSSION

Inspection of the $\Delta \varepsilon_r$ curves allows state some interesting features. (i) The curves of the 1-pentanol + TEGDME system are skewed to high $x_1$ values, and show a rather large minimum (Figure 1). This indicates that the polyether is a very active breaker of the alcohol structure, in such way that the total number of parallel aligned effective dipoles of 1-pentanol, that contribute to the dielectric polarization of the system, decreases upon mixing [26]. A similar behaviour is observed for methanol + TEGDME, or + PEG-250 mixtures (Figure 1). (ii) For 1-pentanol systems, larger $|\Delta\varepsilon_r|$ values are encountered for the mixture with TEGDME than for the DBE solution (Figure 1). That is, the linear monoether is a less active compound when breaking the self-association of the 1-alkanol. (iii) In addition, the $\Delta\varepsilon_r$ curve is much more symmetrical for the DBE mixture, which clearly reveals that effects related to the alcohol self-association are here more important than in the corresponding pentaether system, (iv) For mixtures including TEGDME, $\Delta\varepsilon_r$ (methanol) $> \Delta\varepsilon_r$ (1-pentanol), which can be ascribed to the higher self-association of methanol. For example, in terms of the ERAS model, the equilibrium constant of methanol is 986 while that of 1-pentanol is 153 [5,8]. (v) Interestingly, $\Delta\varepsilon_r$ results for methanol + TEGDME or + PEG-250 systems are rather similar (Figure 1), although it seems that PEG-250 breaks more easily the alcohol self-association. A systematic study on effect of the increasing the number of (CH$_2$CH$_2$O) groups on $\Delta\varepsilon_r$ measurements is currently undertaken. (vi) In the case of octane solutions, $\Delta\varepsilon_r$ (TEGDME) $<$ $\Delta\varepsilon_r$ (DBE) (Figure 2). This merely reveals that the polyether is a more structured liquid than DBE.

The temperature dependence of $\varepsilon_r$ has been examined from the parameter:

$$\alpha = \frac{1}{\varepsilon_r} \frac{d\varepsilon_r}{d(1/T)} \qquad (6)$$

Figure 3 shows that $\alpha$ is positive at any composition, which is the normal behaviour of dipolar liquids. For 1-pentanol + organic solvent mixtures, at $x_1 > 0.4$, $\alpha$ changes in the sequence: TEGDME < DBE < octane. This suggests that the mixture structure of the TEGDME solution changes in less extent at increased temperature values due to the higher ability of this ether to break the self-association of the alcohol at room temperature. The disappearance of the minimum of the $\alpha$ curve for the octane systems at lower 1-pentanol concentrations can be explained in similar terms as such minimum has been ascribed to the existence of cyclic multimers of 1-pentanol at those concentrations [16]. On the other hand, at low $x_1$ values, $\alpha$ (methanol + TEGDME) $<$ $\alpha$ (1-pentanol + TEGDME). This indicates that association effects are more relevant for the methanol system. Finally, for octane solutions, $\alpha$ (TEGDME) $> \alpha$ (DBE), in agreement with the higher structure of TEGDME.

We have calculated the Kirkwood correlation factor, $g_K$, of 1-pentanol or methanol + TEGDME, or PEG-250 mixtures, and of the TEGDME + octane system. Our $\varepsilon_r$ and

$n_D$ measurements were used for the systems with 1-pentanol or octane. For methanol solutions, $\varepsilon_r$ and $n_D$ data were taken from [17-19]. Density data, needed for calculations are available in the literature [27-30]. Along calculations the molar weight of PEG-250 was taken as 283.33 g [31]. For a system containing one polar compound and one non-polar component, $g_K$ can be determined by the Fröhlich equation [12,13,32,33]

$$g_K = \frac{9k_B T \varepsilon_0 (2\varepsilon_r + \varepsilon_1^\infty)^2}{N_A \mu_1^2 x_1 (\varepsilon_1^\infty + 2)^2 (2\varepsilon_r + 1)} [\frac{V_m(\varepsilon_r - 1)}{\varepsilon_r} - \frac{3V_1 x_1 (\varepsilon_1^\infty - 1)}{2\varepsilon_r + \varepsilon_1^\infty} - \frac{3V_2 x_2 (n_{D2}^2 - 1)}{2\varepsilon_r + n_{D2}^2}] \quad (7)$$

where the symbols have their usual meaning [16]. The high frequency relative permittivity of the polar compound 1, $\varepsilon_1^\infty$, was calculated from the Clausius-Mossotti equation [15], adopting the atomic polarization ($P_A$), evaluated using the relation $P_A + P_E = 1.1 P_E$, where $P_E$, the electronic polarization, was calculated by the Lorenz-Lorentz equation using the refractive index for the sodium-D line [12,13,33]. For a binary mixture involving two polar compounds, $g_K$ is given by [15,34,35]:

$$g_K = \frac{9k_B T V_m \varepsilon_0 (\varepsilon_r - \varepsilon_r^\infty)(2\varepsilon_r + \varepsilon_r^\infty)}{N_A \mu^2 \varepsilon_r (\varepsilon_r^\infty + 2)^2} \quad (8)$$

Here, $\varepsilon_r^\infty$ is the high frequency relative permittivity of the system, calculated similarly to $\varepsilon_1^\infty$ (see above) [36,37] and $\mu$ represents the gas phase dipole moment of the solution, estimated from the equation [34]:

$$\mu = x_1 \mu_1 + x_2 \mu_2 \quad (9)$$

where $\mu_i$ stands for the dipole moment in the gas phase of component i: 1.65 D for 1-pentanol [38]; 2,47 D for TEGDME [39], and 2.8 for PEG-250 [40] (1 D = 3.3564 10$^{-30}$ C·m).

It is interesting to compare $g_K$ curves of the 1-pentanol + octane, or + DBE, or + TEGDME systems. We note that, at low alcohol concentrations, $g_K$ (octane) < $g_K$ (DBE) < $g_K$ (TEGDME) (Figure 4), and the typical minimum encountered in the $g_K$ curves of 1-alcohol + alkane mixtures [12-14,33,41] vanishes when the non polar compound is replaced by an ether. This minimum has been ascribed to the existence of cyclic species, mainly tetramers [14], characterized, according to the rules of vector addition, by dipole moments equal to zero. Thus, the observed behaviour confirms that oxaalkanes are more effective breakers of the alcohol structure than alkanes. At higher alcohol concentrations, $g_K$ changes in opposite way to that stated above. If one takes into account that the dipole moment of a linear chain is larger than that of a monomer, such $g_K$ variation indicates that effects related to alcohol self-association decrease in the order: octane > DBE > TEGDME. As conclusion, the polyether is a more active molecule when breaking the linear multimers of alcohol, the dominant species at high alcohol concentrations. The $g_K$ curves for the TEGDME or DBE + octane mixtures are also compared in Figure 4. From these results, it is clear that TEGDME is a more structured liquid than DBE.

It is remarkable that at 303.15 K $g_K$ curves of methanol or 1-pentanol +TEGDME or methanol + PEG-250 systems are rather similar (Figure 5). This suggests that the mixture structure is also similar for such solutions.

Interestingly, the molar excess enthalpies, $H_m^E$, of 1-pentanol systems at equimolar composition and 298.15 K, changes in the sequence: 614 (octane) [42] < 843 (DBE) [43] < 1956 (TEGDME) [44] (all values in J·mol$^{-1}$), which clearly remarks that effects due to the alcohol self-association are stronger in alkane solutions, while $H_m^E$ of the 1-pentanol + TEGDME mixture is essentially determined by dipolar interactions between like molecules. This is supported by the concentration dependence of the $H_m^E$ curves, which are progressively skewed, from lower mole fractions of 1-pentanol for the octane solution [42], to higher alcohol concentrations for the pentaether mixture [44].

Let's now examine the temperature dependence of $g_K$ for 1-pentanol solutions. In the dilute alcohol region, $g_k$ increases with the temperature, while the opposite trend is observed at high alkanol concentrations. Thus, for the octane system [16], $(\Delta g_K / \Delta T)_{298.15}(x_1 = 0.05) = 0.026$ K$^{-1}$; $(\Delta g_K / \Delta T)_{298.15}(x_1 = 0.90) = -0.015$ K$^{-1}$. The values for the corresponding TEGDME mixture are: $(\Delta g_K / \Delta T)_{298.15}(x_1 = 0.05) = 0.001$ K$^{-1}$; $(\Delta g_K / \Delta T)_{298.15}(x_1 = 0.90) = -0.008$ K$^{-1}$. The observed $(\Delta g_K / \Delta T)_{298.15}$ decrease at $x_1 = 0.05$ when octane is replaced by the pentaether merely expresses that a low number of cyclic species can be broken by the ether when temperature is increased. Similarly, the higher $(\Delta g_K / \Delta T)_{298.15}$ value at $x_1 = 0.90$ for this mixture means that a lower number of linear multimers can be broken by the polyether at increased temperatures. That is, oxaalkanes are better breakers of the alcohol structure than alkanes. On the other hand, at equimolar composition: $(\Delta g_K / \Delta T)_{298.15} = -0.013$ K$^{-1}$ (octane) < $-0.006$ K$^{-1}$ (DBE) < $-0.0005$ K$^{-1}$ (TEGDME). This is consistent with the higher values of the excess heat capacities at constant pressure, $C_{p,m}^E$, of 1-alkanol + alkane mixtures compared with those of 1-alkanol + ether solutions. Thus, equimolar composition and 298.15 K, $C_{p,m}^E$ /J·mol$^{-1}$·K$^{-1}$ = 11.7, for ethanol + heptane [45], and is 7.2 for ethanol + methyl butyl ether [46]. For 1-alkanol + polyether mixtures $C_{p,m}^E$ values are lower, which is a characteristic behavior of mixtures where strong dipolar interactions are present [47]. In the case of the 1-propanol + 2,5,8-trioxanonane mixture, $C_{p,m}^E$ = 4.4 J·mol$^{-1}$·K$^{-1}$ [46].

The molar polarization of the mixtures (or polarizability volume), $P_m$, was calculated according to the equation [34]:

$$P_m = \frac{(\varepsilon_r - n_D^2)(2\varepsilon_r + n_D^2)V_m}{9\varepsilon_r} \quad (10)$$

The very different shape of the $P_m$ curves is remarkable (Figures 6,7). We note that $P_m$ of 1-pentanol + octane, or + DBE mixtures increases regularly with the alcohol concentration (Figure 6), which indicates that the alignment of the alcohol dipoles becomes progressively predominant [16]. The $P_m$ curve of the 1-pentanol + TEGDME system shows a minimum at $x_1 \approx 0.6$ (Figure 6). This reveals a decrease of the mixture polarization which can be explained in terms of an increase in the number of alcohol multimers broken by the polyether upon mixing. The larger $P_m$ values of this solution or of the methanol + TEGDME or + PEG-250 systems are due to the higher molar polarizations of polyethers. TEGDME has a $P_m$ value (320 cm$^3$·mol$^{-1}$) close to that of 1-pentanol (339 cm$^3$·mol$^{-1}$), and much higher than those of DBE (57.5 cm$^3$·mol$^{-1}$) or octane (0.7 cm$^3$·mol$^{-1}$). At 303.15 K, $P_m$ of PEG-250 is 402.3 cm$^3$·mol$^{-1}$. Note the large differences between $P_m$ values of the TEGDME or DBE + octane mixtures (Figure 7). Data on $n_D$ have been used also to calculate molar refractions, $R_m$, (Figures 8,9) according to the equation [48,49]:

$$R_m = \frac{(n_D^2 - 1)V_m}{(n_D^2 + 2)} \quad (11)$$

This equation is obtained from the Lorentz-Lorenz equation by replacing the refractive index at infinite frequency by the refractive index at optical frequencies (usually, the sodium D line) [15,48,49]. Figure 8 plots $R_m$ results for some 1-alkanol + organic solvent mixtures. $R_m$ decreases with increased alkanol concentrations, and this indicates that the polarization of the alkane or ethers is higher than that of the alkyl chain of 1-alkanols. For 1-pentanol systems, $R_m$ increases as follows: octane < DBE < TEGDME. Due to $R_m$ is regarded as a measure of dispersion forces within fluids, this means that such forces become more important in the same order as above, that is, when increasing the number of oxygen atoms in the solvent. Consequently, $R_m$ (methanol + PEG-250) > $R_m$ (methanol + TEGDME). The fact that, for TEGDME solutions, $R_m$ (1-pentanol) > $R_m$ (methanol) reveals that dispersion forces are more relevant in the 1-pentanol systems. Comparison of Figures 8 and 9 shows that $R_m$ is higher for the ether + octane mixtures than for the systems including 1-alkanols at large alcohol concentrations, as in this case orientational polarization is much more important. On the other hand, $R_m$ changes linearly with the concentration and does not depend on the temperature. This is the normal behaviour of systems where no complex formation exist [48,50].

## 5. CONCLUSIONS

The properties $\varepsilon_r$ and $n_D$ have been measured at (293.15-303.15) K for the systems 1-pentanol + TEGDME, or TEGMDE + octane. Values of $\Delta\varepsilon_r$ and $g_K$ for 1-alkanol + TEGDME, or + PEG-250 show that polyethers can break the alcohol structure even at high alcohol concentrations, and that DBE is a less active compound when breaking the alcohol self-association. This is supported by $P_m$ values and by the temperature dependence of $\varepsilon_r$. $R_m$ values of 1-pentanol systems reveal that dispersion forces become more important in the order: octane < DBE < TEGDME. Dispersion forces are also more relevant in the 1-pentanol + TEGDME system than in the methanol + TEGDME solution. Comparison of $\varepsilon_r$ and $g_K$ data for TEGDME or DBE + octane mixtures shows than the polyether is a more structured liquid.

**ACKNOWLEDGEMENTS**

The authors gratefully acknowledge the financial support received from the Ministerio de Ciencia e Innovación, under the Project FIS2010-16957. V.A. acknowledges the grant financed jointly by the Junta de Castilla y León and Fondo Social Europeo.

TABLE 1

Properties of pure compounds at temperature $T$: density, $\rho$, relative permittivity, $\varepsilon_r$, and refractive index, $n_D$.

| Compound | T/K | $\rho$ | | $\varepsilon_r$ | | $n_D$ | |
|---|---|---|---|---|---|---|---|
| | | Exp. | Lit. | Exp. | Lit. | Exp. | Lit. |
| 1-pentanol | 293.15 | 0.81474 | 0.8147 [51] | 15.739 | 15.736 [16] | 1.4100 | 1.4100[16] |
| | 298.15 | 0.81106 | 0.81083 [52] | 15.158 | 15.04 [53] | 1.4082 | 1.4080 [52] |
| | | | 0.81107 [54] | | 15.558 [54] | | 1.40767 [53] |
| | | | 0.8116 [27] | | | | |
| | 303.15 | 0.80739 | 0.8073 [51] | 14.579 | 14.565 [16] | 1.4061 | 1.40573 [55] |
| TEGDME | 293.15 | 1.01106 | 1.0111 [56] | 7.939 | 7.90 [18] | 1.4326 | 1.43224 [19] |
| | | | 1.0114 [57] | | | | 1.43268 [57] |
| | 298.15 | 1.00707 | 1.00668 [20] | 7.816 | 7.78 [20] | 1.4306 | 1.43039 [57] |
| | | | 1.0067 [56] | | | | |
| | 303.15 | 1.00249 | 1.0021 [56] | 7.692 | 7.65[18] | 1.4286 | 1.42802 [19] |
| Octane | 293.15 | 0.70277 | 0.70262 [58] | 1.968 | | 1.3976 | 1.3965 [59] |
| | 298.15 | 0.69873 | 0.69850 [52] | 1.961 | 1.96 [53] | 1.3952 | 1.3951 [52] |
| | | | | | | | 1.3947 [59] |
| | 303.15 | 0.69470 | 0.69449 [60] | 1.954 | | 1.3929 | 1.3926 [59] |

TABLE 2
Relative permittivities, $\varepsilon_r$, and the corresponding deviations form the ideal state, $\Delta\varepsilon_r$, at temperature $T$ for the systems 1-pentanol(1) + 2,5,8,11,14-pentaoxapentadecane(2) or 2,5,8,11,14-pentaoxapentadecane(1) + octane(2).

| $x_1$ | $\varepsilon_r$ | $\Delta\varepsilon_r$ | $\varepsilon_r$ | $\Delta\varepsilon_r$ | $\varepsilon_r$ | $\Delta\varepsilon_r$ |
|---|---|---|---|---|---|---|
| | | 1-pentanol(1) + 2,5,8,11,14-pentaoxapentadecane(2) | | | | |
| | $T$ = 293.15 K | | $T$ = 298.15 K | | $T$ = 303.15 K | |
| 0.0505 | 7.990 | −0.148 | 7.863 | −0.140 | 7.737 | −0.131 |
| 0.1002 | 7.970 | −0.374 | 7.843 | −0.355 | 7.715 | −0.335 |
| 0.1509 | 8.043 | −0.523 | 7.911 | −0.496 | 7.779 | −0.467 |
| 0.1919 | 8.041 | −0.714 | 7.906 | −0.679 | 7.771 | −0.642 |
| 0.2458 | 8.132 | −0.885 | 7.993 | −0.838 | 7.854 | −0.790 |
| 0.2910 | 8.150 | −1.100 | 8.009 | −1.041 | 7.866 | −0.984 |
| 0.3384 | 8.271 | −1.236 | 8.119 | −1.174 | 7.968 | −1.109 |
| 0.3986 | 8.344 | −1.513 | 8.186 | −1.436 | 8.029 | −1.357 |
| 0.4554 | 8.523 | −1.690 | 8.350 | −1.607 | 8.180 | −1.521 |
| 0.4930 | 8.615 | −1.848 | 8.438 | −1.755 | 8.260 | −1.662 |
| 0.5496 | 8.857 | −2.008 | 8.662 | −1.909 | 8.469 | −1.808 |
| 0.5931 | 9.027 | −2.170 | 8.819 | −2.065 | 8.615 | −1.955 |
| 0.6455 | 9.353 | −2.272 | 9.122 | −2.165 | 8.896 | −2.052 |
| 0.6979 | 9.710 | −2.379 | 9.454 | −2.269 | 9.203 | −2.154 |
| 0.7424 | 10.138 | −2.376 | 9.853 | −2.270 | 9.577 | −2.155 |
| 0.7959 | 10.730 | −2.337 | 10.413 | −2.230 | 10.103 | −2.117 |
| 0.8372 | 11.351 | −2.179 | 10.992 | −2.087 | 10.643 | −1.986 |
| 0.8679 | 11.894 | −2.002 | 11.506 | −1.918 | 11.123 | −1.829 |
| 0.8987 | 12.519 | −1.766 | 12.102 | −1.688 | 11.691 | −1.605 |
| 0.9225 | 13.119 | −1.483 | 12.665 | −1.423 | 12.217 | −1.358 |
| 0.9488 | 13.878 | −1.090 | 13.383 | −1.050 | 12.896 | −1.003 |
| 0.9793 | 14.929 | −0.489 | 14.377 | −0.479 | 13.834 | −0.462 |
| | | 2,5,8,11,14-pentaoxapentadecane(1) + octane(2) | | | | |
| 0.0330 | 2.107 | −0.124 | 2.098 | −0.121 | 2.088 | −0.118 |
| 0.0659 | 2.254 | −0.234 | 2.242 | −0.228 | 2.230 | −0.222 |
| 0.0998 | 2.413 | −0.334 | 2.398 | −0.325 | 2.382 | −0.318 |
| 0.1330 | 2.577 | −0.417 | 2.558 | −0.407 | 2.538 | −0.399 |
| 0.1645 | 2.736 | −0.488 | 2.715 | −0.476 | 2.693 | −0.465 |

TABLE 2 (continued)

| | | | | | | |
|---|---|---|---|---|---|---|
| 0.2003 | 2.922 | −0.557 | 2.896 | −0.545 | 2.870 | −0.533 |
| 0.2488 | 3.190 | −0.625 | 3.158 | −0.612 | 3.126 | −0.599 |
| 0.3002 | 3.476 | −0.684 | 3.438 | −0.670 | 3.400 | −0.657 |
| 0.3486 | 3.762 | −0.713 | 3.718 | −0.699 | 3.675 | −0.684 |
| 0.3951 | 4.030 | −0.739 | 3.981 | −0.724 | 3.931 | −0.710 |
| 0.4461 | 4.347 | −0.734 | 4.291 | −0.720 | 4.234 | −0.707 |
| 0.5092 | 4.738 | −0.716 | 4.674 | −0.703 | 4.610 | −0.690 |
| 0.5516 | 5.005 | −0.692 | 4.937 | −0.679 | 4.867 | −0.667 |
| 0.6003 | 5.313 | −0.656 | 5.238 | −0.644 | 5.163 | −0.632 |
| 0.6409 | 5.578 | −0.612 | 5.499 | −0.600 | 5.420 | −0.588 |
| 0.6930 | 5.909 | −0.557 | 5.823 | −0.546 | 5.736 | −0.537 |
| 0.7072 | 6.004 | −0.536 | 5.916 | −0.526 | 5.827 | −0.517 |
| 0.7515 | 6.298 | −0.468 | 6.205 | −0.459 | 6.111 | −0.451 |
| 0.8002 | 6.622 | −0.386 | 6.523 | −0.379 | 6.421 | −0.375 |
| 0.8461 | 6.943 | −0.288 | 6.838 | −0.283 | 6.732 | −0.278 |
| 0.8985 | 7.279 | −0.199 | 7.168 | −0.196 | 7.057 | −0.191 |
| 0.9441 | 7.585 | −0.104 | 7.469 | −0.101 | 7.351 | −0.099 |

TABLE 3
Refractive indices, $n_D$, at temperature $T$ for the systems 1-pentanol(1) + 2,5,8,11,14-pentaoxapentadecane(2) or 2,5,8,11,14-pentaoxapentadecane(1) + octane(2).

| $x_1$ | $n_D$ | | |
|---|---|---|---|
| | 1-pentanol(1) + 2,5,8,11,14-pentaoxapentadecane(2) | | |
| | $T$ = 293.15 K | $T$ = 298.15 K | $T$ = 303.15 K |
| 0.0505 | 1.4319 | 1.4300 | 1.4279 |
| 0.1002 | 1.4315 | 1.4294 | 1.4273 |
| 0.1509 | 1.4305 | 1.4286 | 1.4265 |
| 0.1919 | 1.4301 | 1.4281 | 1.4259 |
| 0.2458 | 1.4291 | 1.4272 | 1.4251 |
| 0.2910 | 1.4285 | 1.4265 | 1.4244 |
| 0.3384 | 1.4276 | 1.4256 | 1.4236 |
| 0.3986 | 1.4265 | 1.4246 | 1.4225 |
| 0.4554 | 1.4256 | 1.4235 | 1.4215 |
| 0.4930 | 1.4248 | 1.4229 | 1.4207 |
| 0.5496 | 1.4236 | 1.4217 | 1.4196 |
| 0.5931 | 1.4227 | 1.4207 | 1.4186 |
| 0.6455 | 1.4214 | 1.4194 | 1.4174 |
| 0.6979 | 1.4200 | 1.4181 | 1.4160 |
| 0.7424 | 1.4188 | 1.4168 | 1.4148 |
| 0.7959 | 1.4172 | 1.4153 | 1.4133 |
| 0.8372 | 1.4159 | 1.4141 | 1.4121 |
| 0.8679 | 1.4149 | 1.4131 | 1.4111 |
| 0.8987 | 1.4139 | 1.4121 | 1.4100 |
| 0.9225 | 1.4130 | 1.4112 | 1.4092 |
| 0.9488 | 1.4120 | 1.4102 | 1.4081 |
| 0.9793 | 1.4109 | 1.4090 | 1.4070 |
| | 2,5,8,11,14-pentaoxapentadecane(1) + octane(2) | | |
| 0.0503 | 1.3993 | 1.3969 | 1.3946 |
| 0.1003 | 1.4011 | 1.3987 | 1.3965 |
| 0.1515 | 1.4031 | 1.4007 | 1.3985 |
| 0.2061 | 1.4052 | 1.4028 | 1.4006 |
| 0.2450 | 1.4067 | 1.4043 | 1.4021 |
| 0.3050 | 1.4091 | 1.4067 | 1.4045 |

TABLE 3 (continued)

| | | | |
|---|---|---|---|
| 0.3569 | 1.4111 | 1.4087 | 1.4065 |
| 0.4054 | 1.4127 | 1.4106 | 1.4084 |
| 0.4487 | 1.4144 | 1.4122 | 1.4100 |
| 0.5043 | 1.4163 | 1.4142 | 1.4121 |
| 0.5537 | 1.4182 | 1.4159 | 1.4138 |
| 0.5963 | 1.4195 | 1.4173 | 1.4151 |
| 0.6473 | 1.4213 | 1.4191 | 1.4170 |
| 0.6930 | 1.4228 | 1.4208 | 1.4186 |
| 0.7515 | 1.4248 | 1.4227 | 1.4207 |
| 0.8067 | 1.4268 | 1.4248 | 1.4230 |
| 0.8475 | 1.4281 | 1.4260 | 1.4239 |
| 0.8941 | 1.4295 | 1.4275 | 1.4256 |
| 0.9470 | 1.4310 | 1.4290 | 1.4271 |

TABLE 4

Coefficients $A_i$ and standard deviations, $\sigma$ (equation 5) for representation of the $\Delta\varepsilon_r$ and $n_D$ properties at temperature $T$ for 1-pentanol(1) + 2,5,8,11,14-pentaoxapentadecane(2) or 2,5,8,11,14-pentaoxapentadecane(1) + octane(2) systems by equation 3 ($\Delta\varepsilon_r$) or by equation 4 ($n_D$)

| Property | $T$/K | $A_1$ | $A_2$ | $A_3$ | $A_4$ | $A_5$ | $\sigma$ |
|---|---|---|---|---|---|---|---|
| \multicolumn{8}{c}{1-pentanol(1) + 2,5,8,11,14-pentaoxapentadecane(2)} |
| $\Delta\varepsilon_r$ | 293.15 | −7.45 | −6.73 | −4.82 | −4.51 | −2.58 | 0.019 |
| $n_D$ |  | 1.4322 | −0.0083 | −0.0135 |  |  | 0.0001 |
| $\Delta\varepsilon_r$ | 298.15 | −7.08 | −6.43 | −4.63 | −4.39 | −2.59 | 0.018 |
| $n_D$ |  | 1.4303 | −0.0087 | −0.013 |  |  | 0.0002 |
| $\Delta\varepsilon_r$ | 303.15 | −6.70 | −6.11 | −4.39 | −4.26 | −2.58 | 0.017 |
| $n_D$ |  | 1.4281 | −0.0083 | −0.013 |  |  | 0.0002 |
| \multicolumn{8}{c}{2,5,8,11,14-pentaoxapentadecane(1) + octane(2)} |
| $\Delta\varepsilon_r$ | 293.15 | −2.90 | 0.78 | −0.08 | 0.32 |  | 0.005 |
| $n_D$ |  | 1.3971 | 0.0405 | −0.005 |  |  | 0.0002 |
| $\Delta\varepsilon_r$ | 293.15 | −2.84 | 0.76 | −0.06 | 0.29 |  | 0.004 |
| $n_D$ |  | 1.3947 | 0.041 | −0.005 |  |  | 0.0001 |
| $\Delta\varepsilon_r$ | 303.15 | −2.78 | 0.73 | −0.05 | 0.29 |  | 0.004 |
| $n_D$ |  | 1.3932 | 0.0365 |  |  |  | 0.0003 |

**CAPTION TO FIGURES**

**FIG. 1** $\Delta \varepsilon_r$ at temperature $T$ for some 1-alkanol(1) + ether(2) mixtures. Symbols, experimental values. Full symbols, 1-pentanol(1) + TEGDME(2) mixtures (this work): (▲), $T$ = 293.15 K; (■), $T$ = 298.15 K; (●), $T$ = 303.15 K. Open symbols: (O), 1-pentanol(1) + DBE(2) ($T$ = 298.15K) [16]; (□), methanol(1) + TEGDME(2) ($T$ = 303.15 K) [18]; (Δ), methanol(1) + PEG-250(2) ($T$ = 303.15 K) [17]. Solid lines, smoothed values from Redlich-Kister expansions.

**FIG. 2** $\Delta \varepsilon_r$ at temperature $T$ for ether(1) + octane(2) mixtures. Symbols, experimental values. Full symbols, TEGDME mixtures (this work): (▲), $T$ = 293.15 K; (■), $T$ =298.15 K; (●), $T$ = 303.15 K. Open symbols, DBE system at 298.15 K [16]. Solid lines, smoothed values from Redlich-Kister expansions.

**FIG. 3** $\alpha$ parameter (equation 6) at 298.15 K. Symbols: (●), 1-pentanol(1) + TEGDME(2); (■), TEGDME(1) + octane(2) (this work). Solid curves: (1) 1-pentanol(1) + octane(2); (2), 1-pentanol(1) + DBE(2); (3), DBE(1) + octane(2) [16]. Dashed line, methanol(1) + TEGDME(2) [18].

**FIG. 4** Correlation factor, $g_K$, at 298.15 K for the systems: (■), 1-pentanol(1) + TEGDME(2) (this work); (□), 1-pentanol(1) + octane(2); (O), 1-pentanol(1) + DBE(2) [16]. Dashed lines: (1), TEGDME(1) + octane (this work); (2), DBE(1) + octane(2) [16]

**FIG. 5** Correlation factor, $g_K$, at 303.15 K for 1-alkanol(1) + polyether(2) systems (this work): (●), 1-pentanol(1) + TEGDME(2); (■), methanol(1) + TEGDME(2); (▲), methanol(1) + PEG-250(2).

**FIG. 6** Molar polarization (this work), $P_m$, at temperature $T$ for the systems: 1-pentanol(1) + TEGDME(2) (●), + DBE(2) (□) or + octane(2) (O) at 298.15 K; and for methanol(1) + TEGDME(2) (■) or + PEG-250 (▼) at 303.15 K.

**FIG. 7** Molar polarization, $P_m$, at 298.15 K for ether(1) + octane(2) systems: (●), TEGDME (this work); (O), DBE [16]

**FIG. 8** Molar refraction (this work), $R_m$, at 298.15 K for the systems: 1-pentanol(1) + TEGDME(2) (●), + DBE(2) (■) or + octane(2) (▲) at 298.15 K; and for methanol(1) + TEGDME(2) (O) or + PEG-250 (□) at 303.15 K.

**FIG. 9** Molar refraction, $R_m$, at 298.15 K for ether(1) + octane(2) systems: (●), TEGDME; (O) (this work), DBE [16]

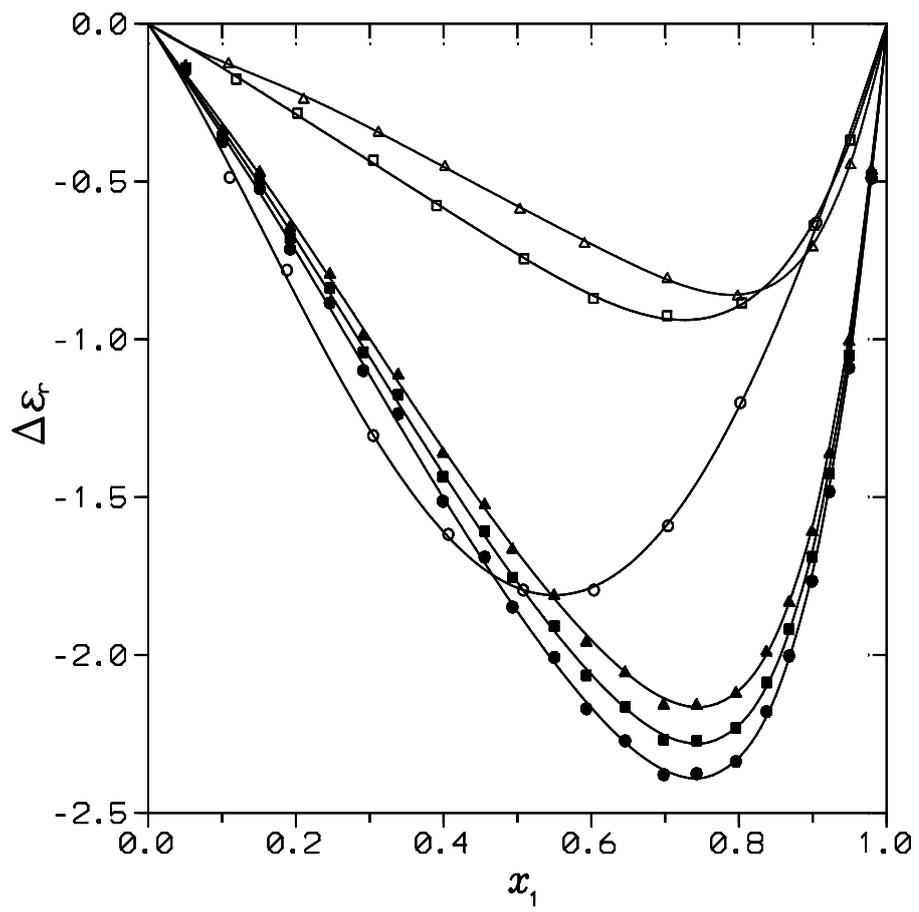

Figure 1

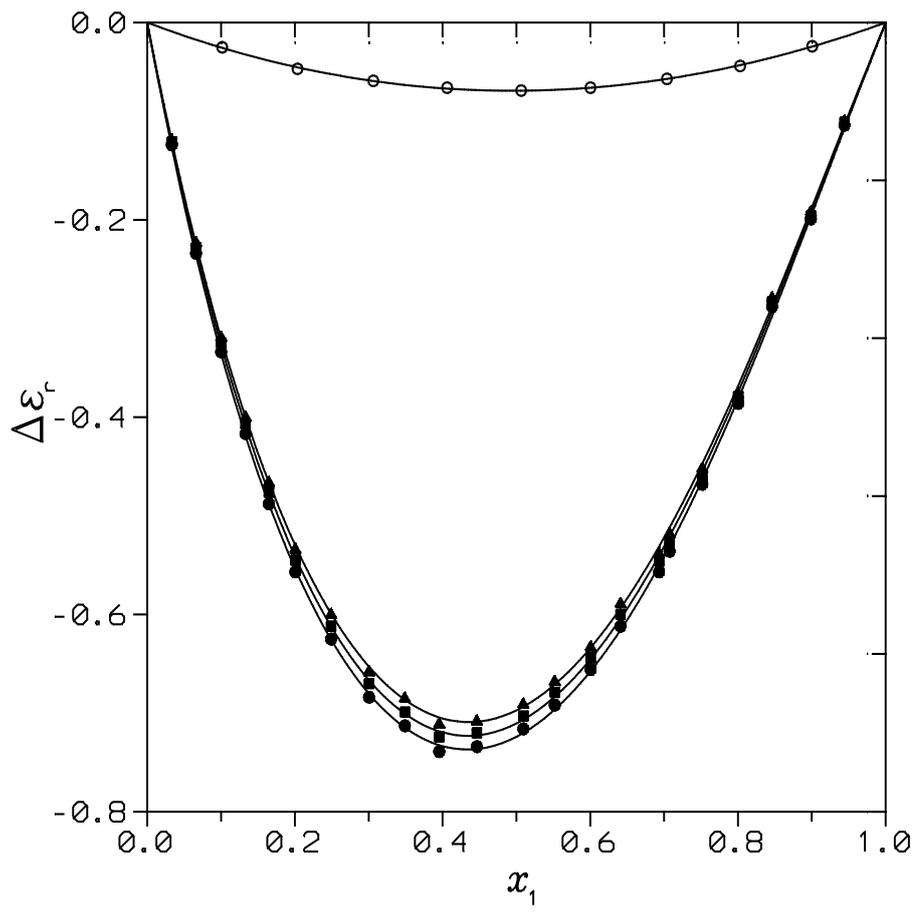

Figure 2

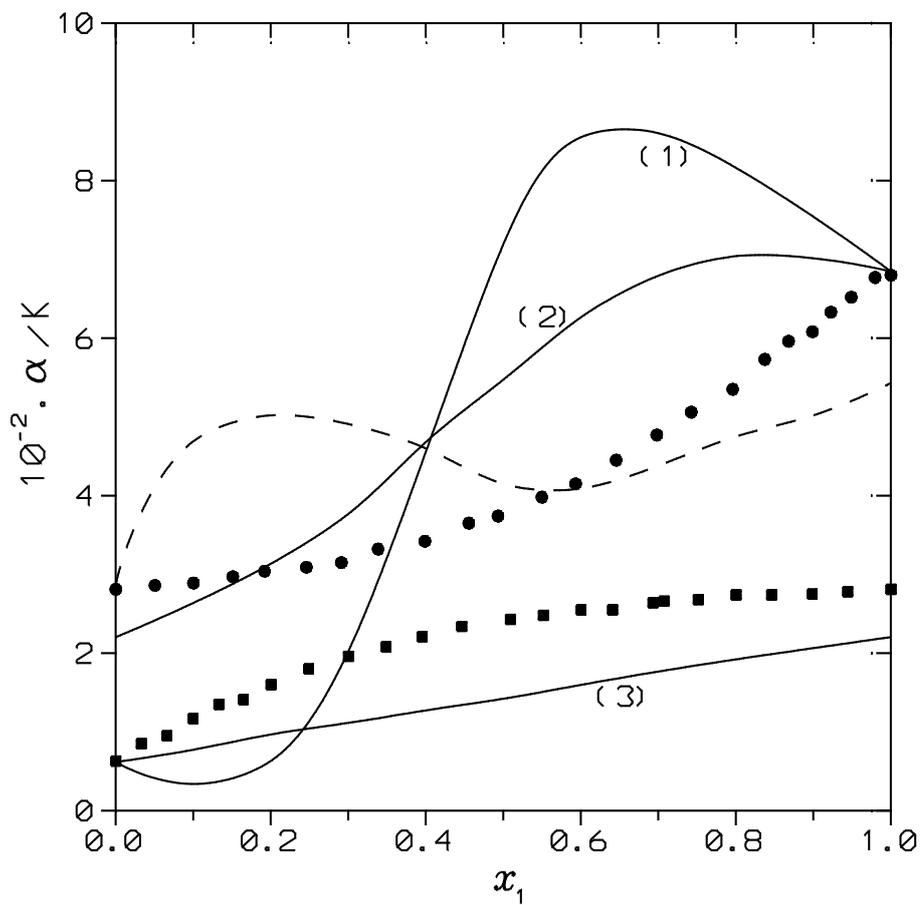

Figure 3

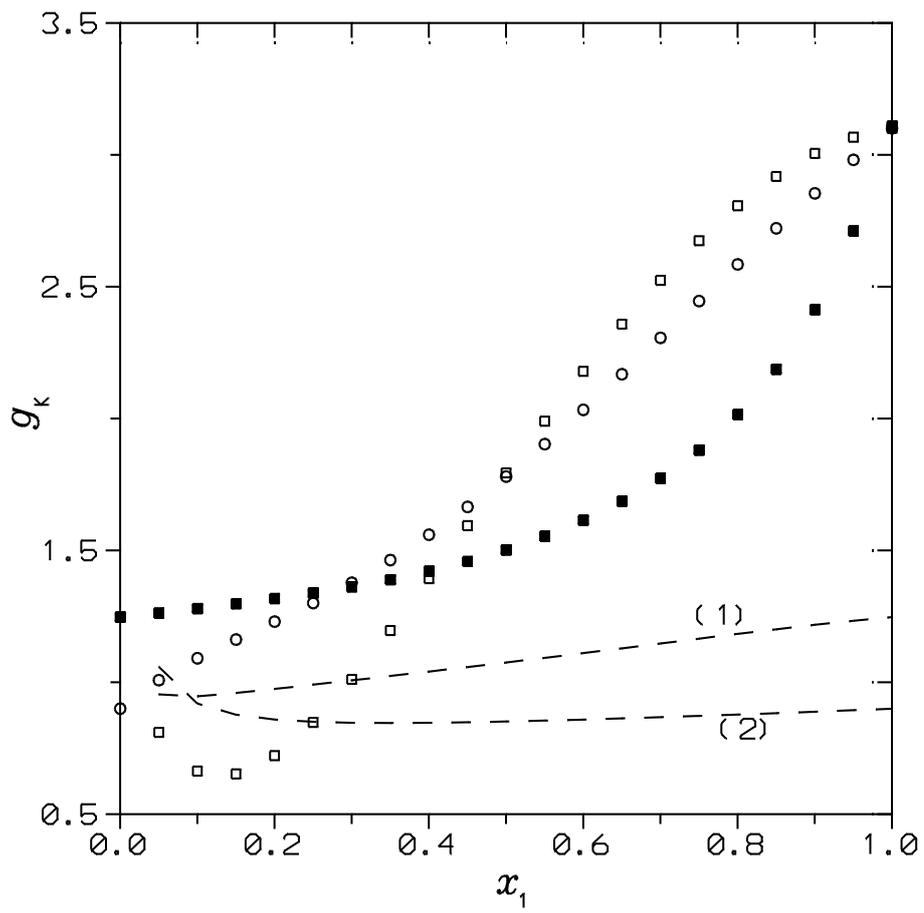

Figure 4

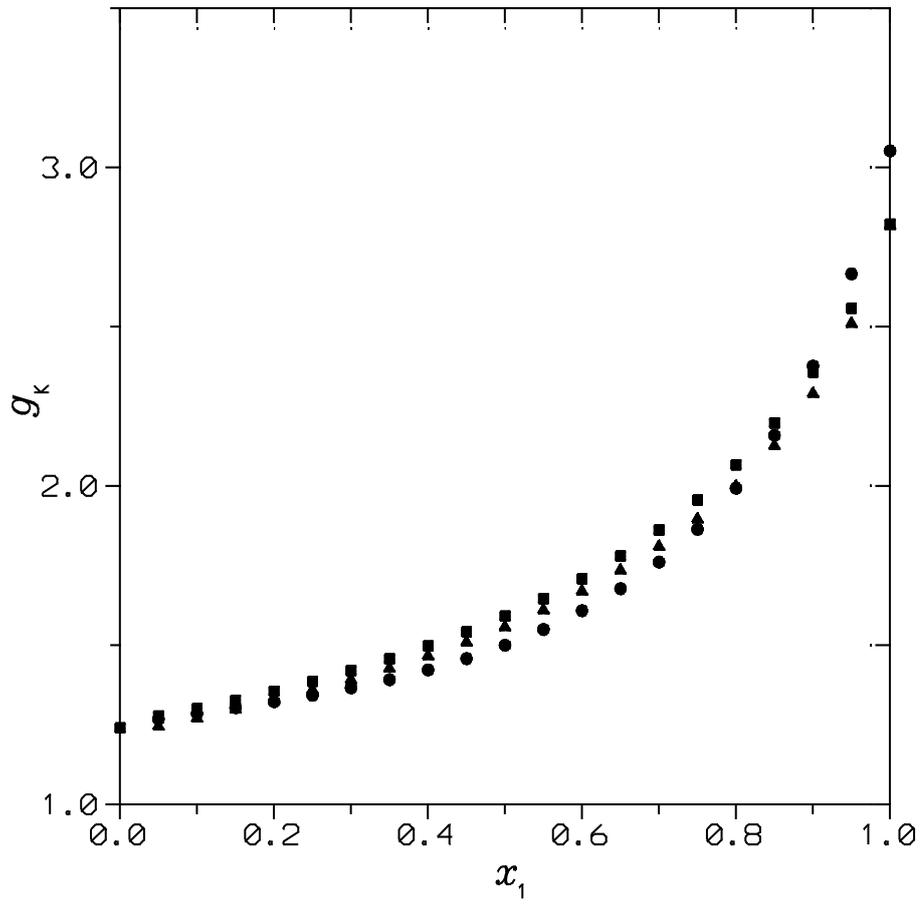

Figure 5

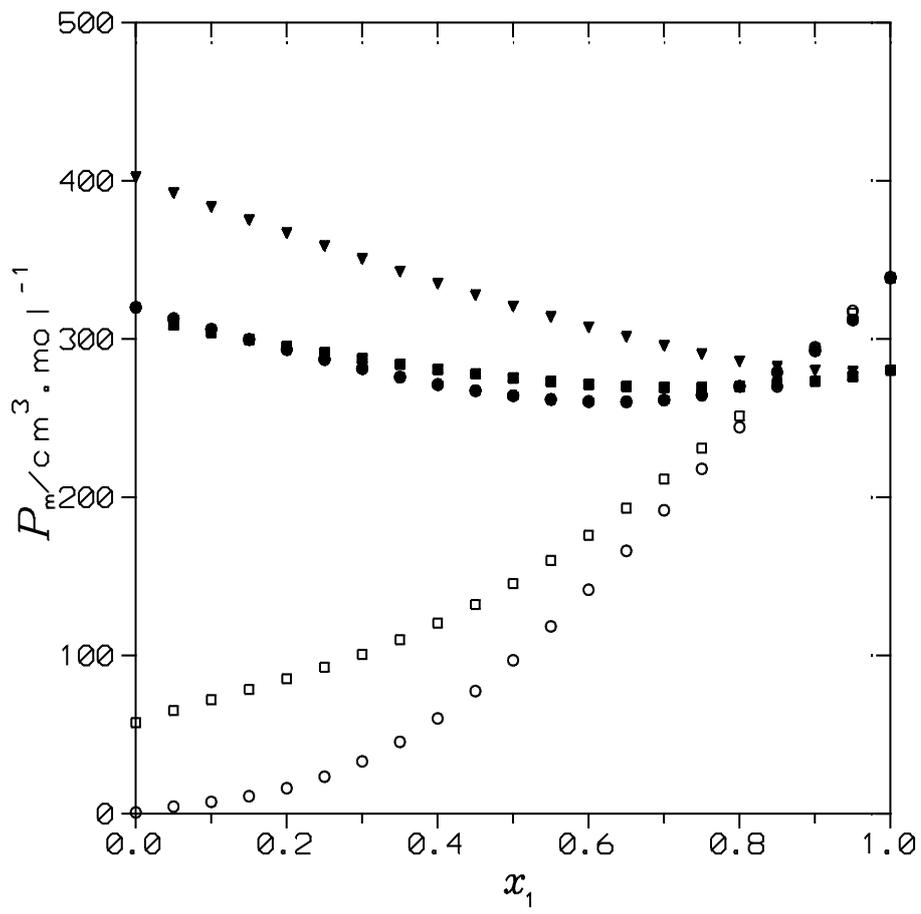

Figure 6

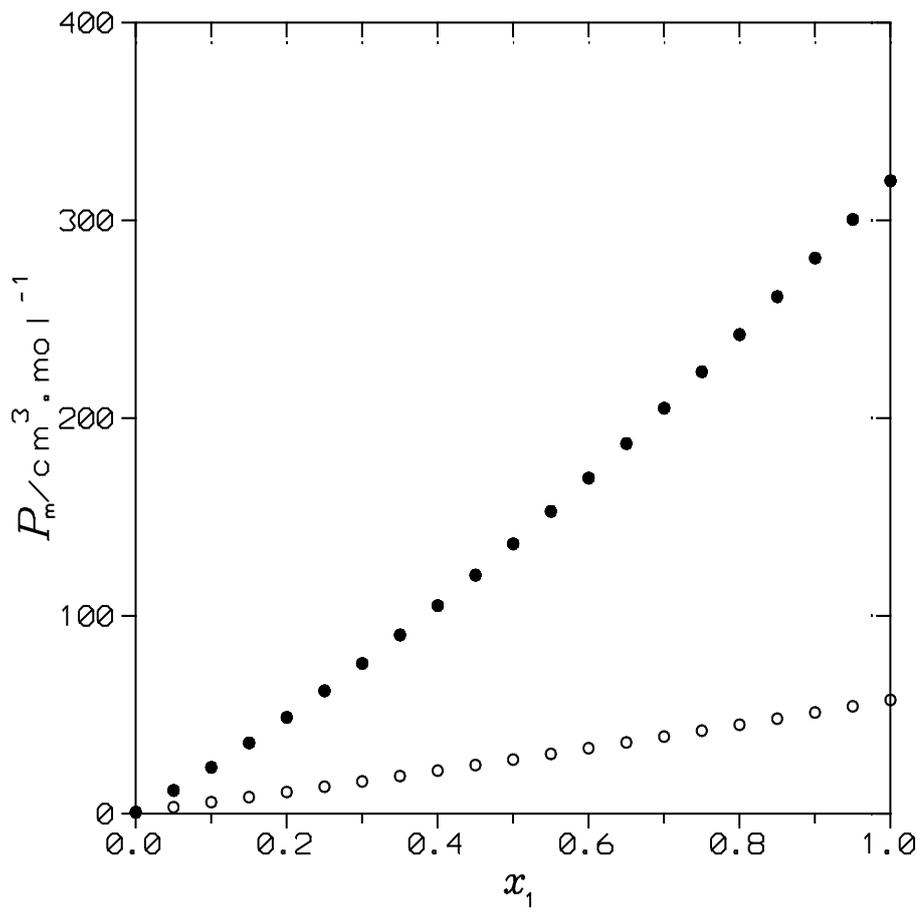

Figure 7

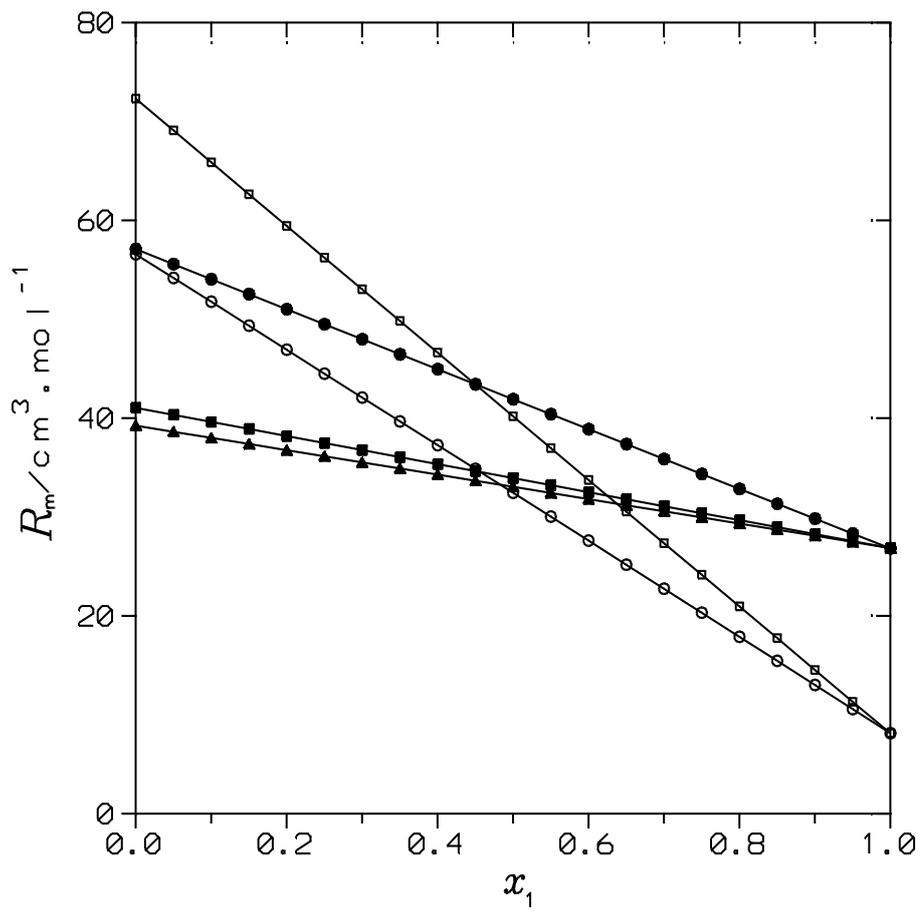

Figure 8

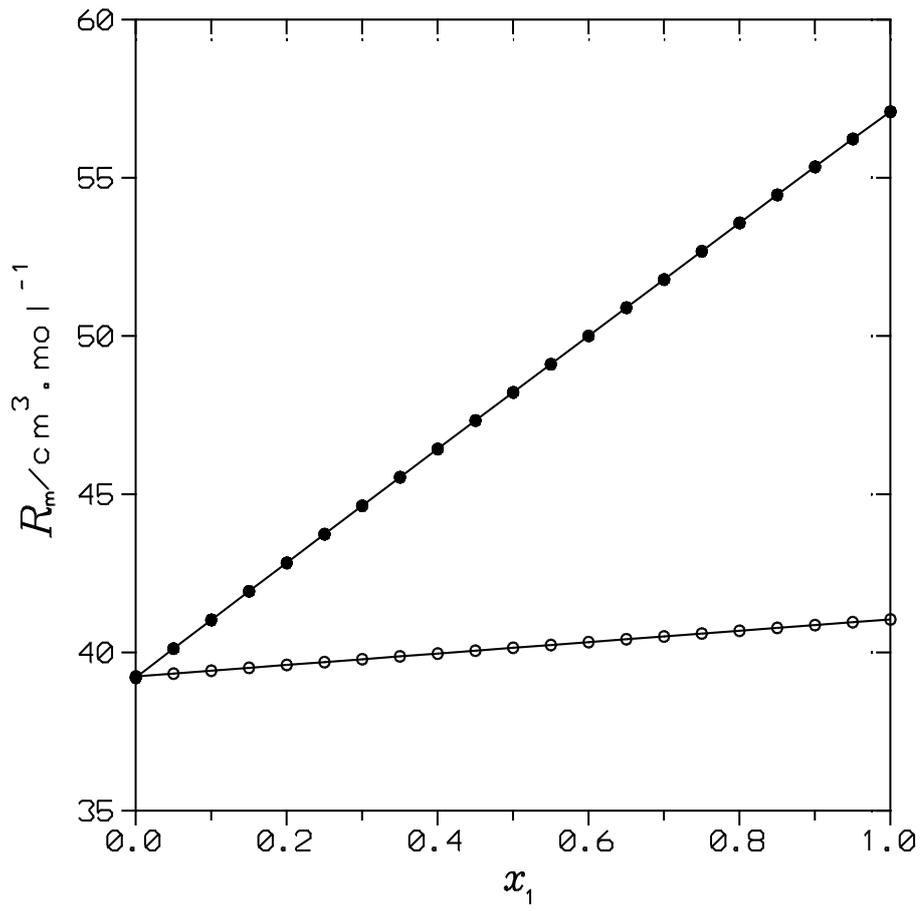

Figure 9